\documentclass[showpacs,preprintnumbers,amsmath,amssymb,superscriptaddress,aps,twocolumn]{revtex4}

\usepackage{amsmath}
\usepackage{amssymb}
\usepackage{amsfonts}
\usepackage{amsthm}
\usepackage{fancyhdr}
\usepackage{fancybox}
\usepackage{graphicx}
\usepackage{natbib}
\usepackage{float}

\renewcommand{\Re}{{\rm Re}}
\renewcommand{\Im}{{\rm Im}}
\makeatletter

\newcommand{\ket}[1]{|{#1}\rangle}
\newcommand{\kb}[1]{|{#1}\rangle\!\langle{#1}|}
\newcommand{\bk}[1]{\langle{#1}|{#1}\rangle}

\newcommand{\bra}[1]{\langle{#1}|}
\newcommand{\bkt}[2]{\langle{#1}|{#2}\rangle}
\newcommand{\Ref}[1]{(\ref{#1})}

\newcommand{\ot}{\otimes}

\newcommand{\av}[1]{\langle#1\rangle}

\begin{document}
\title{Optimal probe wave function of weak-value amplification}
 \author{Yuki Susa}
 \email{susa@th.phys.titech.ac.jp}
 \affiliation{Department of Physics, Tokyo Institute of Technology, Tokyo, Japan}
 \author{Yutaka Shikano}
 \email{yshikano@ims.ac.jp}
 \affiliation{Department of Physics, Tokyo Institute of Technology, Tokyo, Japan}
 \affiliation{Schmid College of Science and Technology, Chapman University, CA, USA}
 \affiliation{Institute for Molecular Science, Okazaki, Aichi, Japan}
 \author{Akio Hosoya}
 \email{ahosoya@th.phys.titech.ac.jp}
 \affiliation{Department of Physics, Tokyo Institute of Technology, Tokyo, Japan}
\date{\today}
\begin{abstract}
The weak measurement proposed by Aharonov and his colleagues extracts information about a physical
quantity of the system by the postselection as shifts of the argument of the probe wave function. The more the
postselected state is orthogonal to the preselected state, the larger the shift determined by the weak value becomes.
Recently, the signal amplification by the weak measurement has been extensively studied. In the present work,
we explicitly obtain the optimal probe wave function and the amplification factor for a given weak value, which
is calculated from the experimental setup. It is shown that the amplification factor has no upper bound, in contrast
to the Gaussian probe wave function, and that the amplified signal is sharp.
\end{abstract}
\maketitle
%%%%%%%%%%%%%%%%%%%%%%
\section{Introduction}
%%%%%%%%%%%%%%%%%%%%%%
Light has been widely used for highly sensitive sensing devices, well-known examples of which are an optical biosensor~\cite{Bio_rev} and a single-atom addressing~\cite{Greiner}. To obtain information from tiny objects by using light, the optical signal should be magnified, and various practical techniques for the signal amplification have been developed, e.g., in Refs.~\cite{Sig_rev1,Sig_rev2}.

As one of the methods of the signal amplification, the weak-value amplification has recently been demonstrated~\cite{Resch}.  
Historically speaking, the idea of this method comes from the {\it weak measurement} initiated by Aharonov et al.~\cite{AAV}. 
The weak measurement was proposed as the time-symmetric quantum measurement~\cite{ABL} almost without destroying a quantum state. 
The measured quantity is called the {\it weak value}, which consists of the observable and the pre- and postselected states. 
Taking the postselection of the system state is the key difference from the conventional quantum measurement. The preselected state corresponds to the state preparation, while the postselected one corresponds to the detection. 
These states together constitute the context of a given weak measurement. 
The formal description of the weak measurement in the von Neumann interaction case was shown in Ref.~\cite{Jozsa}. 
While there are many physical proposals and demonstrations to measure the weak value by the weak measurement reviewed in Ref.~\cite{Shikano}, we shall only consider the von Neumann interaction between the system and the probe in this paper. 

The original idea of the weak measurement was to extract the weak value from the displacement of the probe wavefunction, an example of which is shown in Fig. \ref{Mach}. 
The significance of the weak value in the fundamental quantum mechanics is seen in the reviews in~\cite{AR,AV,AT}. 
However, the purpose of the weak-value amplification is to amplify the shift of the expectation value of the probe position. Therefore, the coupling constant is fixed, and the weak value varies by changing the postselected state. 
By using the weak measurement, the weak-value amplification was demonstrated to experimentally verify the spin Hall effect of light~\cite{Hosten,Pfeifer} and to measure the beam deflection by tilting the mirror~\cite{Dixon,Starling1,Starling2,Turner} and the frequencies through the prism~\cite{Starling3} in the Sagnac interferometer. 
Since the weak value can be arbitrarily large by choosing the almost-orthogonal pre- and postselected states, the weak-value amplification provides us with a new technique for the signal amplification. 
On the other hand, the probability should be very small in the case of the almost-orthogonal pre- and postselected states. 
In order to set the large amplification, we have to repeat the weak measurements many times or, more practically, use an incident beam of sufficient intensity~\cite{Hosten,Dixon,Starling1,Starling2,Turner,Starling3}, e.g., classical light~\cite{Howell,Steinberg}. It is theoretically shown that the small longitudinal phase shifts can
be detected~\cite{Brunner,Li}.
While there is no experimental demonstration for the solid-state system, the charge sensing amplification using the weak value was theoretically proposed~\cite{Gefen}. 
However, Wu and Li theoretically pointed out that the effect of the back action is important in the weak-value amplification on the basis of the second-order calculation~\cite{Wu}. 
The full-order calculation is needed to study the relation between the weak-value amplification and the measurement back action. 
It is numerically shown that there exists an upper bound of the weak-value amplification in the context of the cross-Kerr effect~\cite{Feizpour}. 
Analytically, there also exists an upper bound under the assumption that the probe wave function is Gaussian and the observable $\hat{A}$ satisfies $\hat{A}^2 = 1$~\cite{Zhu,Nakamura,Koike}. 
In many optical applications, the $\hat{A}^{2}=1$ condition holds.
Here, the intriguing question is raised whether the weak-value amplification has the upper bound or not in general. 

In this paper, we will give an analytical expression for the optimal probe wave function in momentum space, which maximizes the factor of the weak-value amplification. 
In our present task, we fix the coupling constant and use the weak value of the observable $\hat{A}$ such that $\hat{A}^2 = 1$ is determined from a given experimental setup.
We shall explicitly show that this optimal case has no upper bound in the amplification. 
Furthermore, the signal after the quantum measurement is sharp around the final probe position.

The rest of the present paper is organized as follows. In Sec.~\ref{wm_sec}, we recapitulate the weak measurement in the weak- and general- coupling cases under the assumption that the probe wave function is Gaussian and the observable $\hat{A}$ satisfies $\hat{A}^2 = 1$. In Sec.~\ref{main_sec}, we show the analytical expression for the optimal probe wave function in the momentum space and examine its properties. Our main result is derived by the variational method to maximize the probe shift given the weak value and the coupling constant in Sec.~\ref{derivation_sec}. Section~\ref{conc_sec} is devoted to the summary and the discussion.
In Appendix A, detailed calculations are shown. An explicit computation of the weak value is demonstrated in the Mach-Zehnder interferometer.
Throughout this paper, we use the unit $\hbar=1$.

%%%%%%%%%%%%%%%%%%%%%%
\section{Weak Measurement:A Brief Review}
%%%%%%%%%%%%%%%%%%%%%%
\label{wm_sec}

In this section, we briefly review the weak measurement proposed by Aharonov et al.~\cite{AAV}.
The weak measurement is characterized by the pre- and postselections of the system state.
We prepare the initial state $\ket{\phi_{i}}$ of the system and $\ket{\psi_{i}}$ of the probe. 
After a certain interaction between the system and the probe, we postselect a system state $\ket{\phi_{f}}$ and obtain information about a physical quantity $\hat{A}$ from the probe wave function by the weak value
\begin{eqnarray}
\label{weak_value}
A_{w}:=\frac{\bra{\phi_{f}}\hat{A}\ket{\phi_{i}}}{\bkt{\phi_{f}}{\phi_{i}}},
\end{eqnarray}
which  can generally be a complex number.
More precisely, the shifts of the position and momentum in the probe wave function are given by the real and imaginary parts of the weak value $A_{w}$, respectively. 
We can easily see from Eq. \Ref{weak_value} that when $\ket{\phi_{i}}$ and $\ket{\phi_{f}}$ are almost orthogonal, the absolute value of the weak value can be arbitrarily large.
This leads to the weak-value amplification, as we will explain below.
As a trade-off, the probability of obtaining a postselected state that is almost orthogonal to the preselected state is very small. 
To make the large probe shift definite,  the weak measurement should be performed many times.

For the weak measurement, the coupling interaction is taken to be the standard von Neumann Hamiltonian, 
\begin{eqnarray}
H=g\delta(t-t_{0})\hat{A}\ot \hat{p},
\end{eqnarray}
where $g$ is a coupling constant and $\hat{p}$ is the probe momentum operator conjugate to the position operator $\hat{q}$. 
We have taken the interaction to be impulsive at time $t=t_{0}$ for simplicity. 
The time evolution operator becomes $\displaystyle e^{-ig\hat{A} \ot \hat{p}}$. 
After postselection, the probe state becomes 
\begin{eqnarray}
\ket{\psi_{f}}=\bra{\phi_{f}}e^{-ig\hat{A}\ot \hat{p}}\ket{\phi_{i}}\ket{\psi_{i}}.
\end{eqnarray}
We denote the expectation values of the initial and final probe positions as
\begin{eqnarray}
\av{\hat{q}}_{i}:=\frac{\bra{\psi_{i}}\hat{q}\ket{\psi_{i}}}{\bk{\psi_{i}}}, \ \ \ \av{\hat{q}}_{f}:=\frac{\bra{\psi_{f}}\hat{q}\ket{\psi_{f}}}{\bk{\psi_{f}}}.
\end{eqnarray}
The shift of the expectation value of the position is defined by
\begin{eqnarray}
\label{qshifts}
\Delta\av{\hat{q}}:=\av{\hat{q}}_{f}-\av{\hat{q}}_{i}.
\end{eqnarray}
Similarly, we define $\av{\hat{p}}_{i}$, $\av{\hat{p}}_{f}$, and $\Delta \av{\hat{p}}$ by replacing $\hat{q}$ with $\hat{p}$ in the above equations.
Here, we write $\tilde{\xi}_{i}(q):=\bkt{q}{\psi_{i}}$ and $\xi_{i}(p):=\bkt{p}{\psi_{i}}$ as  the initial probe wave functions in the position and momentum spaces, respectively.

To see how the weak value emerges in theory, first consider the weak-coupling case following the original work~\cite{AAV}.
The probe state after the post selection becomes 
\begin{eqnarray}
\ket{\psi_{f}} &=&\bra{\phi_{f}}e^{-ig\hat{A}\ot \hat{p}}\ket{\phi_{i}}\ket{\psi_{i}} \nonumber \\
&=&\bra{\phi_{f}}\left[ 1-ig\hat{A}\ot \hat{p}\right]\ket{\phi_{i}}\ket{\psi_{i}}+O(g^{2}) \nonumber \\
&=&\bkt{\phi_{f}}{\phi_{i}}\left[ 1-igA_{w}\hat{p}\right]\ket{\psi_{i}}+O(g^{2}) \nonumber \\
&=&\bkt{\phi_{f}}{\phi_{i}}e^{-igA_{w}\hat{p}}\ket{\psi_{i}}+O(g^{2})
\end{eqnarray}
for $g |A_{w}|\ll1$.
We assume that the initial probe wave function is Gaussian:
\begin{eqnarray}
\label{gaussian}
\tilde{\xi}_{i}(q) =\left( \frac{2W^{2}}{\pi}\right)^{1/4}e^{-W^{2}q^{2}},
\end{eqnarray}
where $W^{-2}$ is the variance.
Equation \Ref{gaussian} gives $\av{\hat{q}}_{i}=\av{\hat{p}}_{i}=0$.
The probe wave function in the position space after the postselection becomes
\begin{eqnarray}
\bkt{q}{\psi_{f}} &\approx& \bkt{\phi_{f}}{\phi_{i}}e^{-igA_{w}\hat{p}}\bkt{q}{\psi_{i}} \nonumber \\
&=&\bkt{\phi_{f}}{\phi_{i}} e^{ -igA_{w} \left(-i\frac{\partial}{\partial q}\right) }\tilde{\xi_{i}}(q) \nonumber \\
&=&\bkt{\phi_{f}}{\phi_{i}}\left( \frac{2W^{2}}{\pi}\right)^{1/4} \nonumber \\
&&\times e^{-W^{2}(q-g A_{w})^{2}+(g A_{w}W)^{2} },
\end{eqnarray}
and therefore, its absolute value squared is given by
\begin{eqnarray}
|\bkt{q}{\psi_{f}}|^{2} &\approx& |\bkt{\phi_{f}}{\phi_{i}}|^{2}\left( \frac{2W^{2}}{\pi}\right)^{1/2} \nonumber \\
&&\times e^{ -2W^{2}(q-g\Re A_{w})^{2}+2(g\Re A_{w}W)^{2}}.
\end{eqnarray}
Thus, we obtain the probe position shift $\Delta\av{\hat{q}}$ [Eq. \Ref{qshifts}] as
\begin{eqnarray}
\label{q_linear}
\Delta\av{\hat{q}}=\av{\hat{q}}_{f}=\frac{\int dq\ q |\bkt{q}{\psi_{f}}|^{2}}{\int dq |\bkt{q}{\psi_{f}}|^{2}}=g\Re A_{w},
\end{eqnarray}
which is proportional to the real part of the weak value.
By a similar calculation, we also obtain the shift of the expectation value of the probe momentum as 
\begin{eqnarray}
\label{p_linear}
\Delta \av{\hat{p}}=\av{\hat{p}}_{f}=2gW^{2}\Im A_{w},
\end{eqnarray}
which is proportional to the imaginary part of the weak value.
From Eqs. \Ref{q_linear} and \Ref{p_linear}, we can extract the weak value, and we can see that, as the weak value increases the probe position shift is amplified.
This effect is called the weak-value amplification. 
It is emphasized that the first-order approximation in $g$ and $|A_{w}|$ is used in the above calculation.

Next, we look at the exact case for an arbitrary coupling constant $g$. 
We assume that $\hat{A}$ satisfies the property $\hat{A}^{2}=1$. 
In this case, the probe state after postselection is calculated as
\begin{eqnarray}
\ket{\psi_{f}}&=&\bra{\phi_{f}}e^{-ig\hat{A}\ot \hat{p}}\ket{\phi_{i}}\ket{\psi_{i}} \nonumber \\
&=&\bra{\phi_{f}}\left[\cos g\hat{p} - i\hat{A}\sin g\hat{p}\right]\ket{\phi_{i}}\ket{\psi_{i}} \nonumber \\
\label{p-function}
&=&\bkt{\phi_{f}}{\phi_{i}} \left[\cos g\hat{p} - iA_{w}\sin g\hat{p}\right]\ket{\psi_{i}} \nonumber \\
&=&\bkt{\phi_{f}}{\phi_{i}} B(\hat{p})\ket{\psi_{i}}. 
\end{eqnarray}
Here, we have defined
\begin{eqnarray}
\label{B_def}
B(\hat{p}):= \cos g\hat{p} - iA_{w}\sin g\hat{p}
\end{eqnarray}
for later convenience.
Choosing the Gaussian form for the initial wave function \Ref{gaussian}, we obtain the shifts of the expectation values of the position and momentum of the probe as 
\begin{eqnarray}
\label{gauss_exact}
\Delta \av{\hat{q}}= \frac{g\Re A_{w}}{1+\frac{1}{2}(1-|A_{w}|^{2}) \left( e^{-2g^{2}W^{2}} -1\right)},\nonumber \\
\\
\Delta\av{\hat{p}} = \frac{2gW^{2}\Im A_{w}e^{-2g^{2}W^{2}}}{1+\frac{1}{2}(1-|A_{w}|^{2}) \left( e^{-2g^{2}W^{2} }-1\right)}, \nonumber
\end{eqnarray}
respectively~\cite{Nakamura,Zhu}.
We can extract the weak value $A_{w}$ from the shifts $\Delta \av{\hat{q}}$ and $\Delta \av{\hat{p}}$ in Eq. \Ref{gauss_exact}. 
The term $(1-|A_{w}|^{2})[\exp \left[-2g^{2}W^{2} \right] -1]/2$ in the denominator of Eq. \Ref{gauss_exact} is the cause of the upper bound of the amplification in the Gaussian-wave profile of the finite size $1/W$. The term in the denominator exhibits the measurement back action, which comes from the full-order evaluation of $B(\hat{p})\ket{\psi_{i}}$~\cite{Nakamura,Zhu}.

\begin{figure}[t]
 \begin{center}
  \includegraphics[width=85mm]{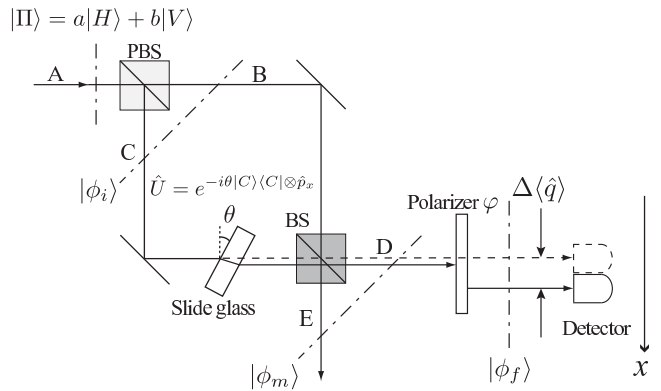}
 \end{center}
 \caption{Mach-Zehnder interferometer with a thin slide glass and a polarizer. The signal comes from port A. The signal detection is taken in at port D. The first and second beam splitters are PBS and the BS, respectively. In path C, the slide glass is inserted. The polarizer, which plays the role of postselection, is placed before the detector in port D. The correspondence to the weak measurement is explained in the text. The pre- and postselected states as well as the intermediate state are explicitly given in Appendix B.}
 \label{Mach}
\end{figure}

A simple example of the weak measurement is illustrated by the Mach-Zehnder interferometer with a thin slide glass and a polarizer in Fig. \ref{Mach} by replacing the first beam splitter with a polarizing beam splitter (PBS).
The system consists of a path state in the basis of $\ket{B}$ and $\ket{C}$ and a polarization state in the basis of $\ket{H}$ and $\ket{V}$. 
Also, the probe is the displacement of the optical axis in port D in the $x$ direction, which is caused by the tilted slide glass, as Fig. 1 indicates.
The interaction in the weak  measurement is introduced by the slide glass tilted by a small angle $\theta$ inserted in path C.
The real part of the weak value can be extracted by the shift $\Delta \av{\hat{q}}$, as explained in Eq. \Ref{q_linear}.
The polarization state $\ket{\Pi}$ is injected from path A.
The initial state $\ket{\phi_{i}}$ and the intermediate state $\ket{\phi_{m}}$ are indicated in Fig. 1.
The tunable angle  $\varphi$ of the polarizer  set in port D controls the postselected state $\bra{\phi_{f}}$.
This setup exemplifies the two aspects of the weak measurement in general. 
First, as frequently emphasized in the literature~\cite{AAV,AR,AV,AT}, we can take information about the weak value in port D from the weak measurement while keeping almost intact the initial polarization state $\ket{\Pi}$ in port E. 
Second, the amplification of the small signal, which is produced by the slight tilt of the slide glass, can be realized by the large shift $\Delta \av{\hat{q}}$ of the optical axis tuned by the polarizer. 
We summarize in Appendix \ref{M_Z} the explicit expressions for the states $\ket{\phi_{i}}$, $\ket{\phi_{m}}$, and $\ket{\phi_{f}}$ and the corresponding weak value.

In the subsequent sections, we will consider general probe wave functions other than the Gaussian wave function and look for the optimal one to obtain the maximum shift $\Delta \av{\hat{q}}$.

%%%%%%%%%%%%%%%%%%%%%%
\section{Main Result}
%%%%%%%%%%%%%%%%%%%%%%
\label{main_sec}
In what follows, we fix a specific experimental setup,
that is, the given coupling constant and the chosen pre- and postselected states, so that we can calculate the weak value before the experiment.
Under this situation, we show the optimal probe wave function, which gives the
maximum shift, and consider its implications and properties in this
section.

The optimal probe wave function in the momentum space is obtained as
\begin{eqnarray}
\label{optxi}
\xi_{i}(p) &=&\sqrt{\frac{g|\Re A_{w}|}{\pi}}\frac{\exp \left[ -i\frac{g(|A_{w}|^{2}+1)}{2\Re A_{w}} p\right]}{\cos gp - iA_{w}\sin gp} \nonumber \\
&=&\sqrt{\frac{g|\Re A_{w}|}{\pi}} B^{-1}(p) \exp\left[ -i\av{\hat{q}}_{f}p\right]
\end{eqnarray}
when $\hat{A}^{2}=1$ and $\Re A_{w} \neq 0$, and the support of the function is $-\pi/2g \leq p \leq \pi/2g$.
The optimal probe wave function gives $\av{\hat{q}}_{i}=0$ and the maximum shift of the expectation value of the probe position as
\begin{eqnarray}
\label{max_shift}
\Delta\av{\hat{q}}=\av{\hat{q}}_{f}=\frac{g(|A_{w}|^{2}+1)}{2\Re A_{w}}.
\end{eqnarray}
We emphasize that the maximum shift is given only by the weak value $A_{w}$ and has no upper bound as $|A_{w}|^{2}$ becomes large. 
On the other hand, as we can see from Eq. \Ref{gauss_exact}, the shifts given by the Gaussian probe wave function have the upper bound because of the back action, as explained before. 
The back-action factor is canceled out by $B^{-1}(p)$ in the expression for the optimal probe wave function, \Ref{optxi}, and therefore we have understood the reason why the amplification has no upper bound.

From Eqs. \Ref{p-function} and \Ref{optxi}, we obtain the final probe wave function in the momentum space as
\begin{eqnarray}
\label{initial_probe_position}
\xi_{f}(p):=\frac{\bkt{p}{\psi_{f}}}{\sqrt{\bk{\psi_{f}}}}=\sqrt{\frac{g}{\pi}}e^{-i\av{\hat{q}}_{f} p},
\end{eqnarray}
where the support of the function is also $-\pi/2g\leq p\leq \pi/2g$.
Performing the inverse Fourier transform, we obtain the final probe wave function in the position space as
\begin{eqnarray}
\label{final_probe_position}
\tilde{\xi}_{f}(q):=\frac{\bkt{q}{\psi_{f}}}{\sqrt{\bk{\psi_{f}}}}=\frac{\sqrt{2g}}{\pi}\frac{\sin \left[\frac{\pi}{2g}(q-\av{\hat{q}}_{f}) \right]}{q-\av{\hat{q}}_{f}},
\end{eqnarray}
where the position $q$ of the probe takes the discrete value $q=2gn$, with $n$ being an integer because of the boundary condition in the momentum space.
Figure 2 displays the initial and final probe wavefunction of the position and momentum spaces in the optimal case. 
Wavefunction \Ref{final_probe_position} is sharp in the neighborhood of $q=\av{\hat{q}}_{f}$, with the width $O(g)$ for a small coupling constant $g \ll 1$ keeping $\av{\hat{q}}_{f}$ finite in the case that interests us most.

As a practical remark on realizing the optimal wave function, we consider a smoothing of the discontinuous  optimal probe wavefunction \Ref{optxi} at the boundary of the finite support. For example, the smoothing function is chosen as 
\begin{equation}
	\xi_{s,i} (p) := \begin{cases}
		\xi_i ( \pi / 2g  ) e^{- (p-\pi /2g )s} & {\rm for} \ p > \pi /2g, \\
		\xi_i ( p ) & {\rm for} \ - \pi / 2g \leq p \leq \pi /2g, \\
		\xi_i ( - \pi / 2g ) e^{(p+\pi /2g) s} & {\rm for} \ p < - \pi /2g \\
		\end{cases}
\end{equation}
with smoothing parameter $s (> 0)$. This smoothing function satisfies  $\lim_{s \to \infty} \xi_{s, i} (p) = \xi (p)$. Then, we can show by explicit calculation that the probe shift by the smoothed
function can be arbitrarily close to the optimal one \Ref{max_shift} for a large $s$. This result for this particular example of smoothing convinces us  that a suitable smoothing does not  drastically change  the wave function or the shift. 
As for the wavefunction this generally holds~\cite{Lebesgue}.

It is interesting to point out that the shift has a lower bound given by
\begin{eqnarray}
|\Delta \av{\hat{q}}|&=&\frac{g}{2}\left( |\Re A_{w}|+\frac{(|\Im A_{w}|^{2}+1)}{|\Re A_{w|}} \right) \nonumber \\
&\geq& g\sqrt{(\Im A_{w})^{2}+1 } \geq g.
\end{eqnarray}
The minimum $|\Delta \av{\hat{q}}|=g$ is attained when $\Re A_{w}=\pm 1$, $\Im A_{w}=0$, and therefore the postselected state coincides with one of the eigenstates of the observable $\hat{A}$.
In this particular case, the weak measurement becomes the projective measurement of the system, and the unitary operator $e^{\mp ig \hat{p}}$ gives the shift operator by $\mp g$ to the probe position. 
It is interesting to note that the weak measurement with the optimal probe wavefunction always amplifies the signal more than the projective measurement.
We also remark that
\begin{eqnarray}
\av{\hat{q}}_{f} |\bkt{\phi_{f}}{\phi_{i}}|\rightarrow \frac{g}{2}
\end{eqnarray}
as the post-selected state $\ket{\phi_{f}}$ approaches the state orthogonal to the preselected state $\ket{\phi_{i}}$. 
From this limit, it is possible to obtain the coupling constant $g$ by extrapolation.

\begin{figure}[t]
\begin{center}
  \includegraphics[width=40mm]{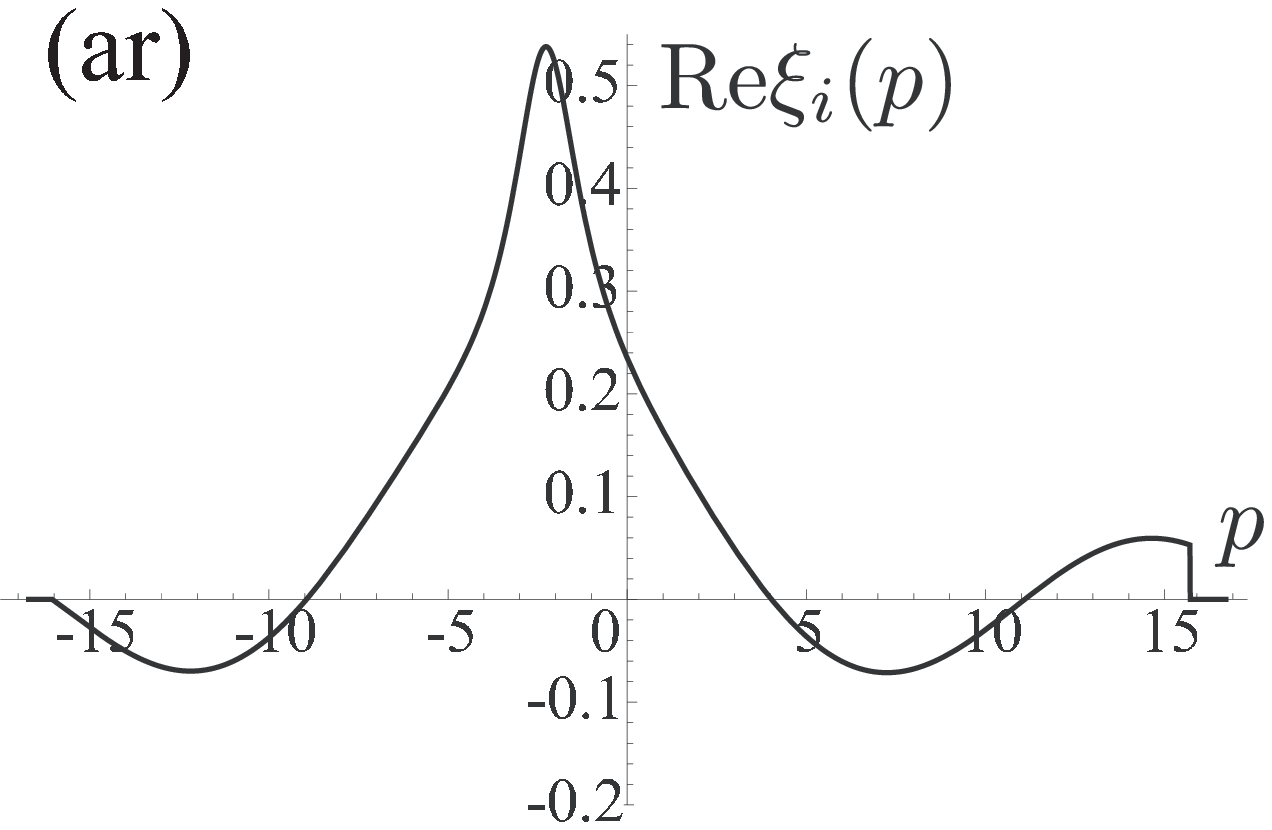}
  \hspace{5pt}
  \includegraphics[width=40mm]{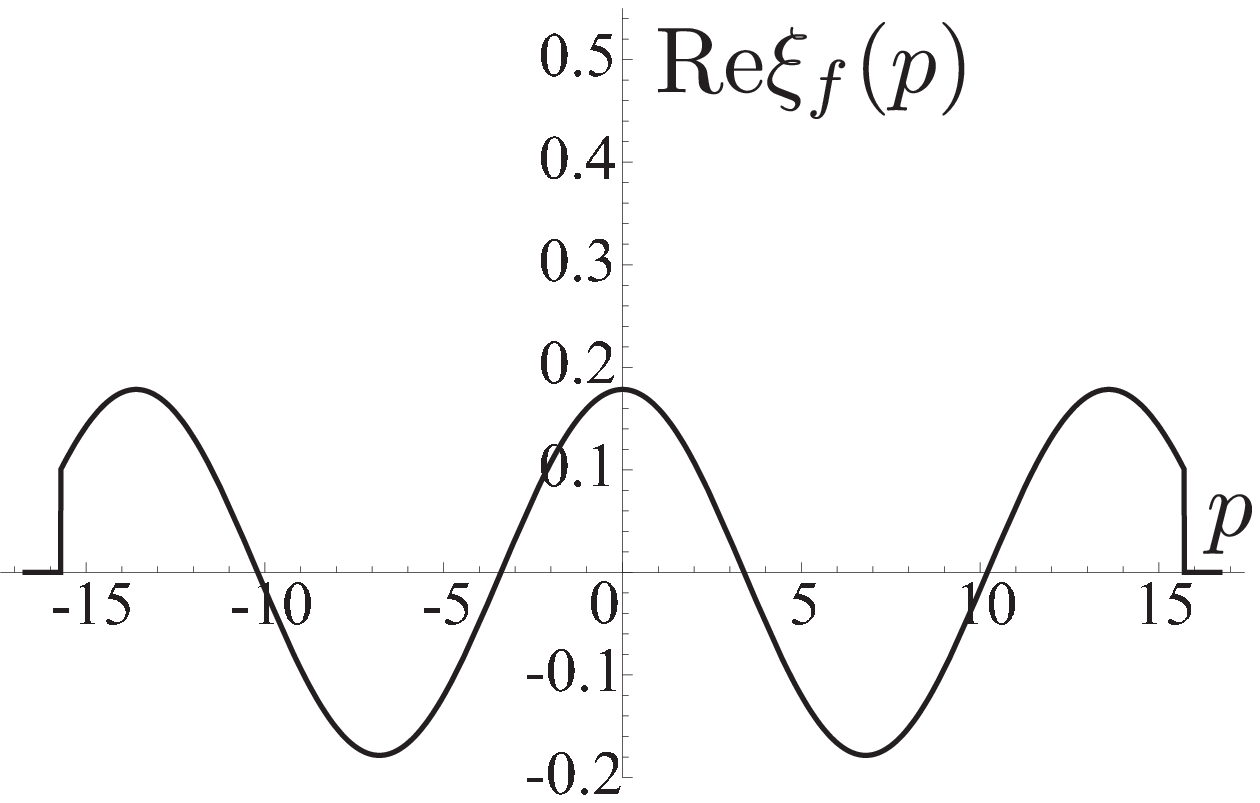}\\
\vspace{8pt}
  \includegraphics[width=40mm]{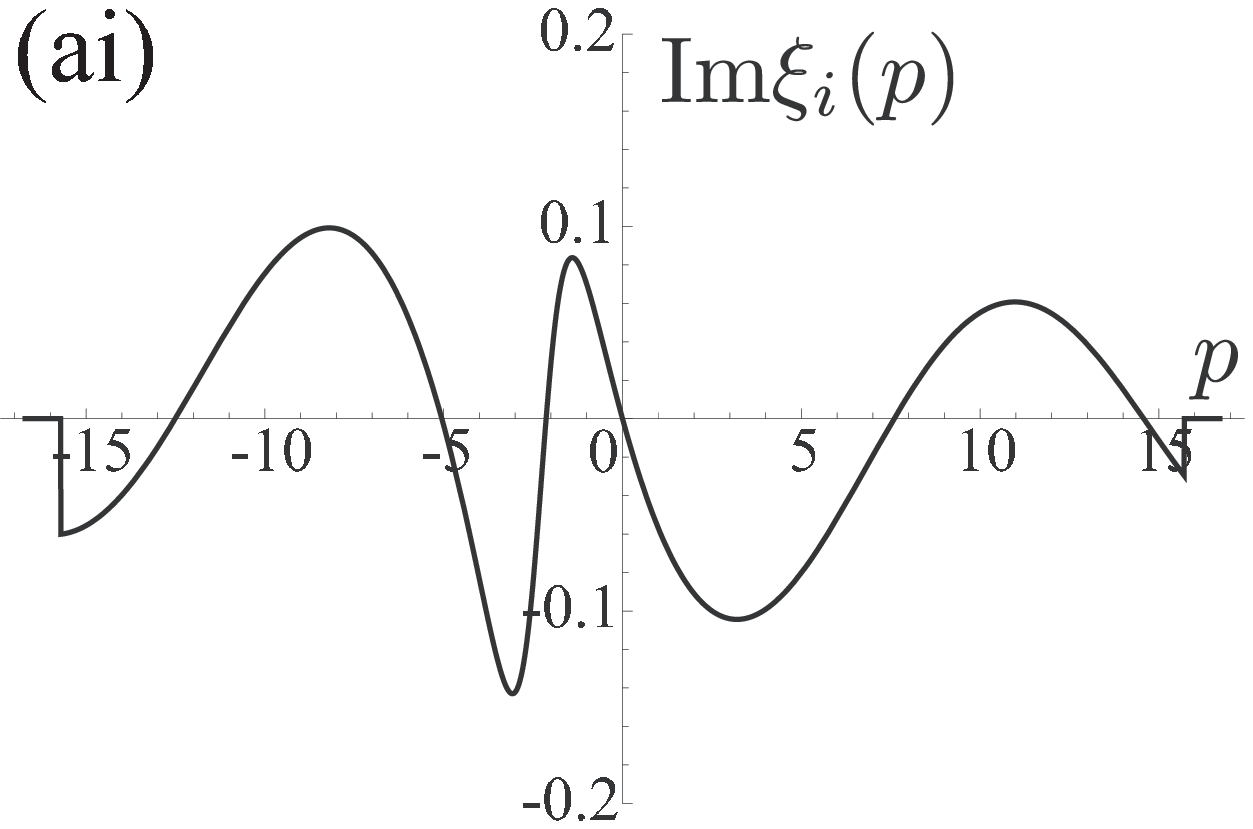}
  \hspace{5pt}
  \includegraphics[width=40mm]{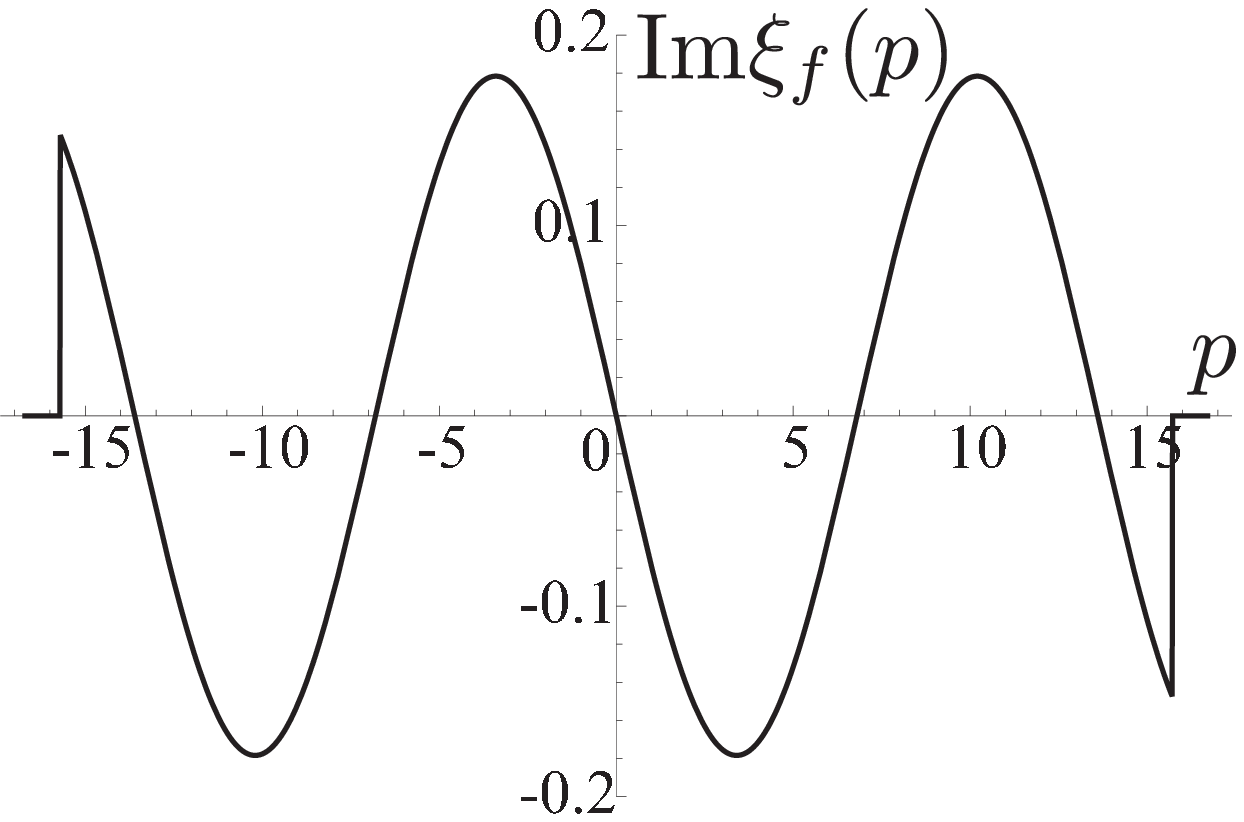}\\
\vspace{8pt}
  \includegraphics[width=40mm]{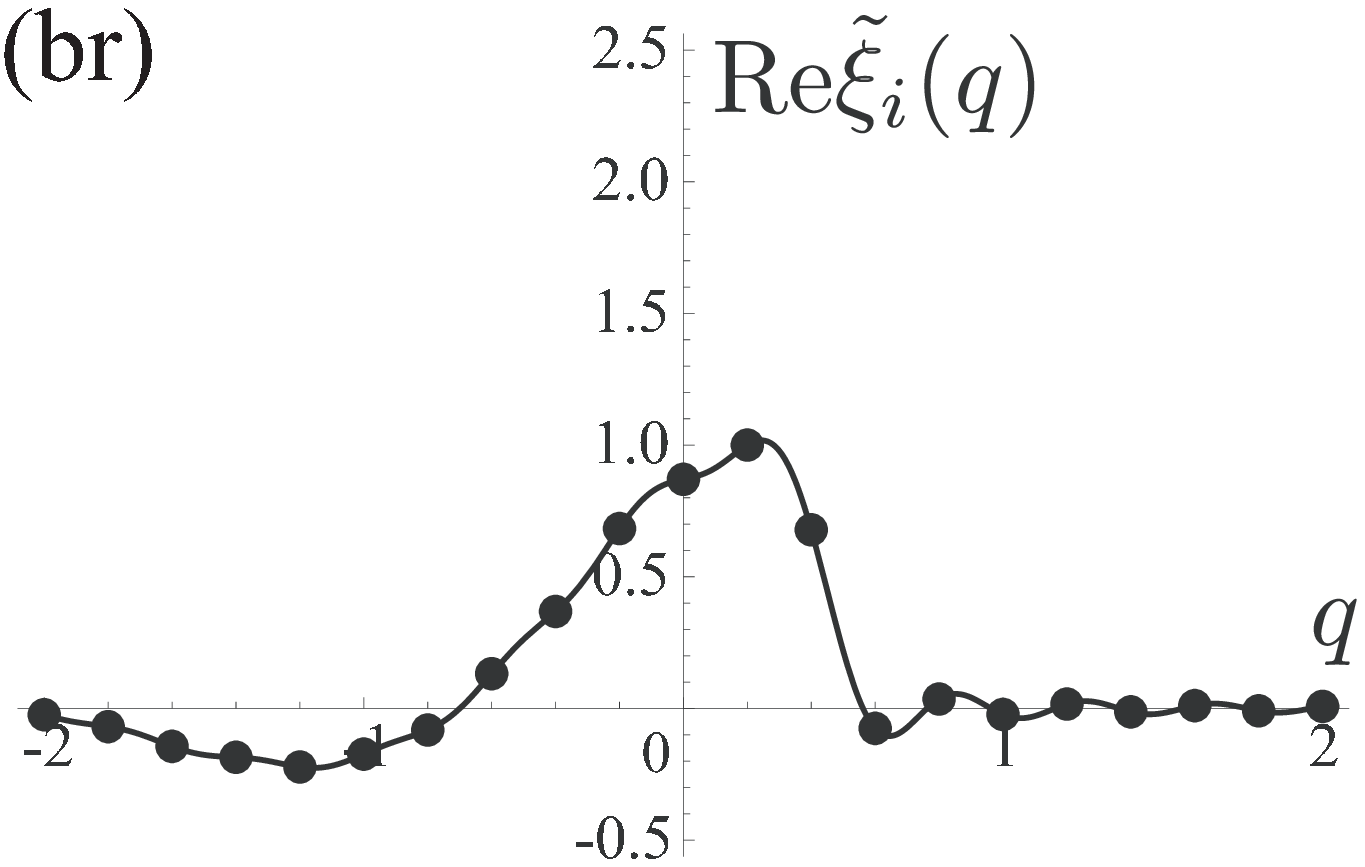}
  \hspace{5pt}
  \includegraphics[width=40mm]{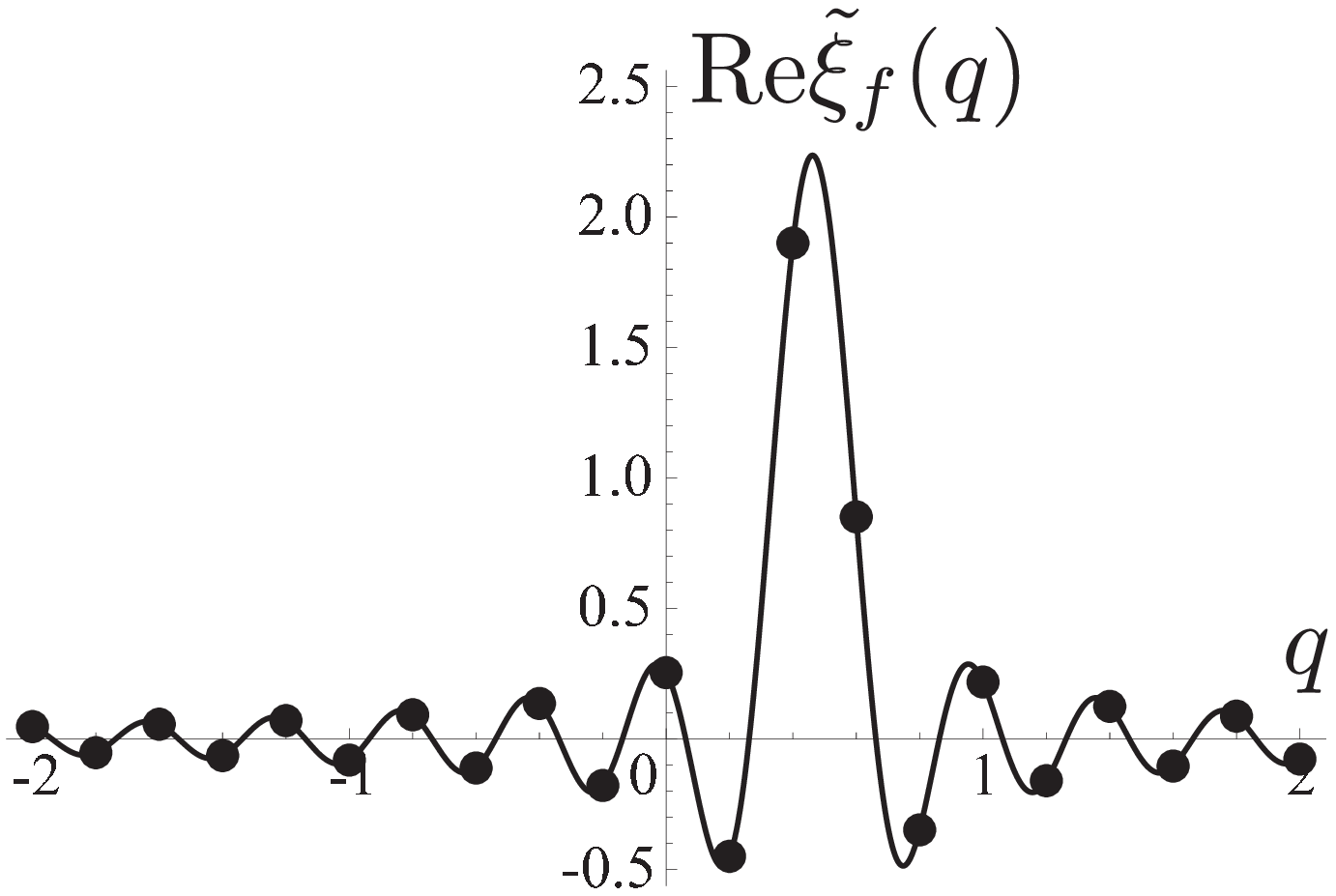}\\
\vspace{8pt}
  \includegraphics[width=40mm]{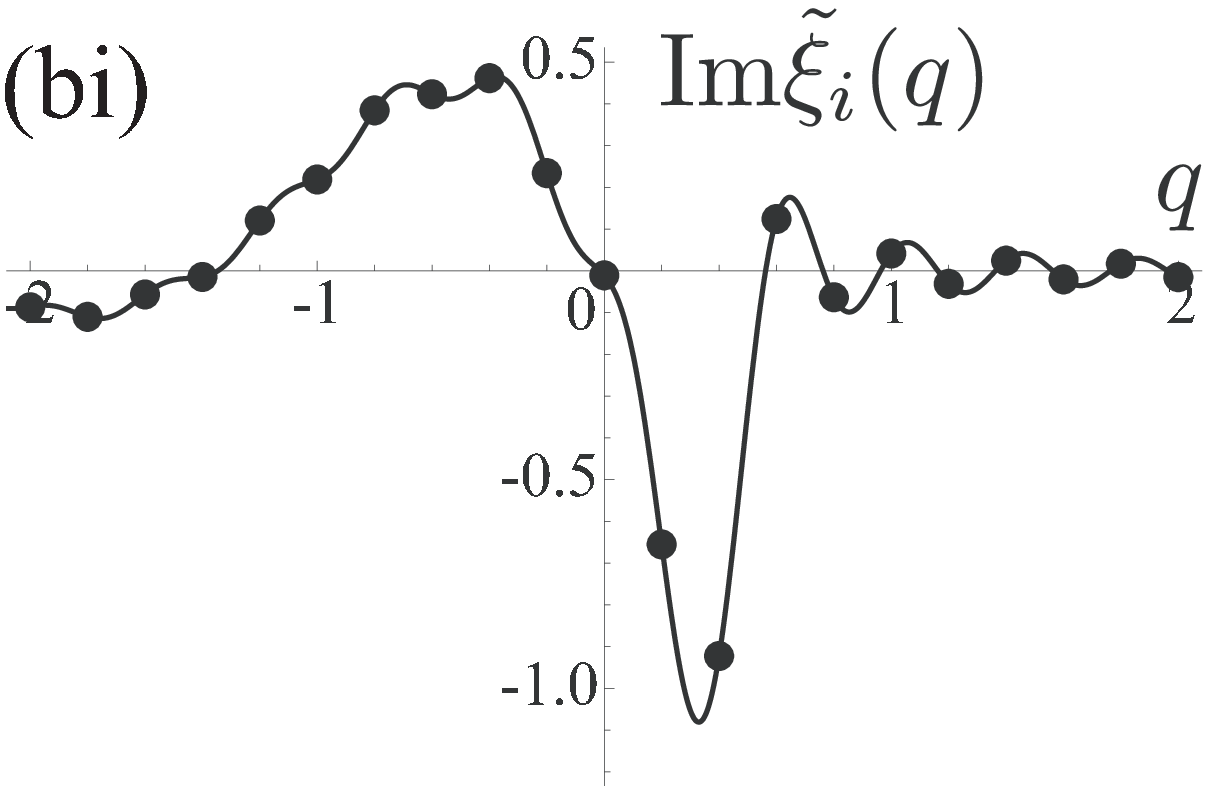}
  \hspace{5pt}
  \includegraphics[width=40mm]{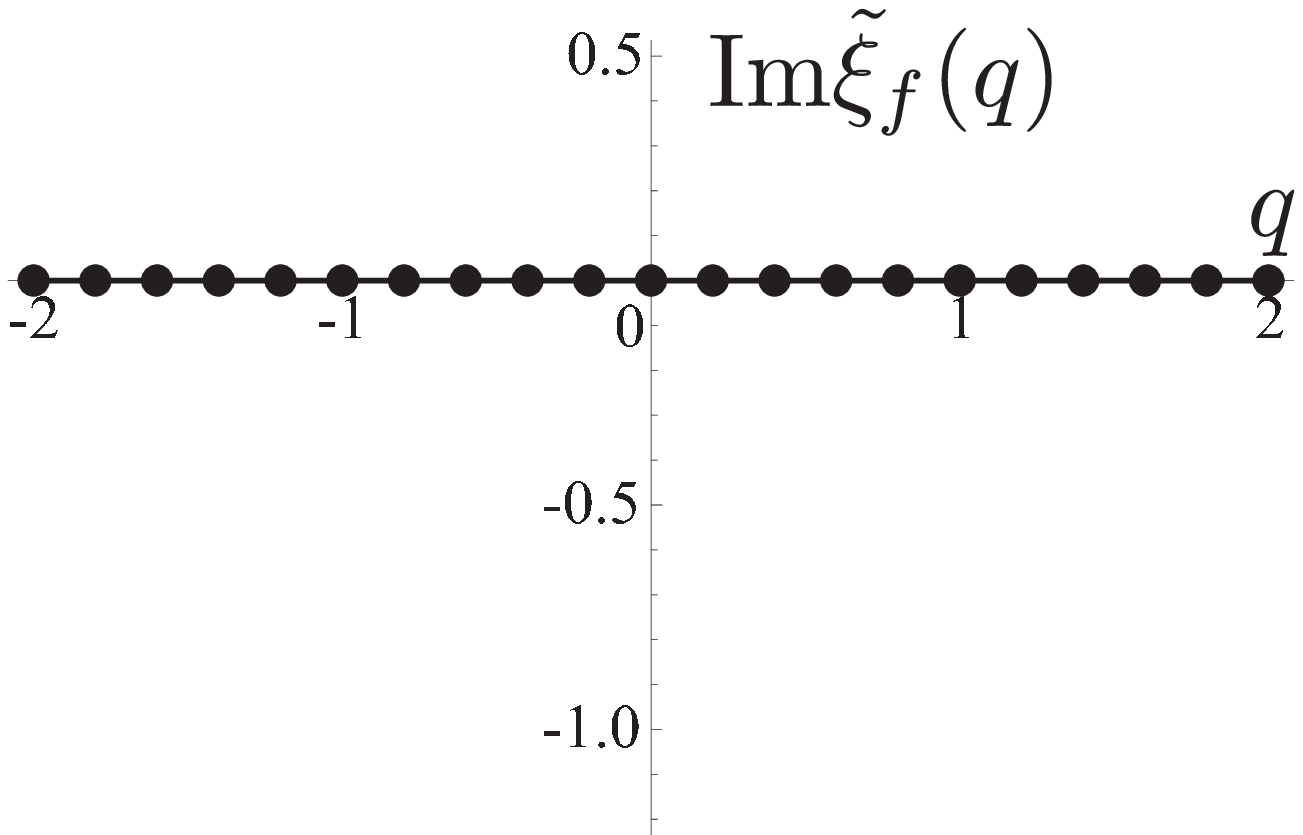}
 \end{center}
 \caption{The initial and the final probe wavefunctions in the optimal case. The left and right sides display the initial and the final probe wave functions, respectively. We set following the parameters: the coupling constant $g = 0.1$ and the weak value $A_{w} = \sqrt{3} + 2\sqrt{3}i$. The plots are the real (ar) and the imaginary (ai) parts of the probe wavefunctions in the momentum space. We also depict the real (br) and the imaginary (bi) parts in the position space. It is noted that the probe wavefunction in the position space is discrete, as the dots indicate.}
 \label{Imxiq}
\end{figure}

%%%%%%%%%%%%%%%%%%%%%%
\section{DERIVATION OF THE MAIN RESULT}
%%%%%%%%%%%%%%%%%%%%%%
\label{derivation_sec}

In this section, we derive the optimal probe wavefunction to obtain the maximum shift using the Lagrange multiplier method.
We consider the probe wave function in the momentum space.
To obtain $\xi_{i}(p)=\bkt{p}{\psi_{i}}$, which gives an extremal value of $\Delta \av{\hat{q}}=\av{\hat{q}}_{f}-\av{\hat{q}}_{i}$, we set a Lagrangian as
\begin{eqnarray}
L[\xi_{i}(p),\xi_{i}^{\ast}(p),\lambda]:=\av{\hat{q}}_{f}-\lambda\left(\int dp |\xi_{i}(p)|^{2}-1 \right),
\end{eqnarray}
where $\lambda$ is a Lagrange multiplier and the constraint condition is the normalization condition for $\xi_{i}(p)$. 
We can set $\av{\hat{q}}_{i}=0$ for convenience, as we will justify subsequently. 
The expectation value of the final probe position $\av{\hat{q}}_{f}$ becomes
\begin{eqnarray}
\label{q_final_probe}
\av{\hat{q}}_{f}&=&\frac{\int dp \bkt{\psi_{f}}{p} \left( i\frac{\partial}{\partial p}\right) \bkt{p}{\psi_{f}}}{\int dp \bkt{\psi_{f}}{p}\bkt{p}{\psi_{f}}} \nonumber \\
&=&\frac{i\int dp [B(p)\xi_{i}(p)]^{\ast} [B(p)\xi_{i}(p)]^{\prime}}{\int dp |B(p)\xi_{i}(p)|^{2}}.
\end{eqnarray}
$B(p)$ is defined in Eq. \Ref{B_def}. 
Varying the Lagrangian $L$ with respect to $\lambda$, we reproduce the normalization condition for $\xi_{i}(p)$ as
\begin{eqnarray}
\label{nomal}
0=\frac{\delta L}{\delta \lambda}=\int dp |\xi_{i}(p)|^{2}-1. 
\end{eqnarray}
Varying $L$ with respect to $\xi_{i}^{\ast}(p)$, we get
\begin{eqnarray}
0&=&\frac{\delta L}{\delta \xi_{i}^{\ast}} \nonumber \\
&=&\frac{i[B^{\ast}(p)B^{\prime}(p)\xi_{i}(p)+|B(p)|^{2}\xi_{i}^{\prime}(p) ]-\av{\hat{q}}_{f} |B(p)|^{2}\xi_{i}(p)}{\int dp |B(p)\xi_{i}(p)|^{2}}\nonumber \\
&&-\lambda \xi_{i}(p).
\end{eqnarray}
This implies
\begin{eqnarray}
\label{logxi}
\frac{\xi_{i}^{\prime}(p)}{\xi_{i}(p)}&=&-\frac{B^{\prime}(p)}{B(p)}-i\left( \av{\hat{q}}_{f}+\lambda |B(p)|^{-2}\int dp |B(p)\xi_{i}(p)|^{2}  \right). \nonumber \\
\end{eqnarray}
Substituting Eq. \Ref{logxi} into Eq. \Ref {q_final_probe}, we find the Lagrange multiplier $\lambda=0$.
Then performing the indefinite integration over $p$ in Eq. \Ref{logxi}, we obtain the probe wave function as
\begin{eqnarray}
\label{xi}
\xi_{i}(p) = C B^{-1}(p) \exp\left[-i\av{\hat{q}}_{f} p \right],
\end{eqnarray}
where $C$ is the normalization constant given by
\begin{eqnarray}
\label{C_0}
|C|^{2}&=&\left(\int dp |B(p)|^{-2}\right)^{-1}
\end{eqnarray}
from Eq. \Ref{nomal}. The expectation value of the final probe position $\av{\hat{q}}_{f}$ is to be determined below.

Hereafter, we evaluate the normalization factor $C$ and the shift of the expectation value of the position $\Delta \av{\hat{q}}$. 
To obtain the shift $\Delta \av{\hat{q}}$, we substitute Eqs.  \Ref{xi} and \Ref{BBprime} into the expectation value of the initial probe position $\av{\hat{q}}_{i}$:
\begin{eqnarray}
\label{qi}
\av{\hat{q}}_{i}&=&\frac{\int dp \bkt{\psi_{i}}{p}\left(i\frac{\partial}{\partial p} \right)\bkt{p}{\psi_{i}}}{\int dp \bkt{\psi_{i}}{p}\bkt{p}{\psi_{i}}} \nonumber \\
&=&i\frac{\int dp \xi_{i}^{\ast}(p)\xi_{i}^{\prime}(p)}{\int dp |\xi_{i}(p)|^{2}}\nonumber \\
&=&\av{\hat{q}}_{f}-i|C|^{2}\int dp\frac{B^{\ast}(p)B^{\prime}(p)}{|B(p)|^{4}} \nonumber \\
&=& \av{\hat{q}}_{f} -g\Re A_{w}|C|^{2}\int dp |B(p)|^{-4} \nonumber \\
&&-\frac{i}{2}|C|^{2}\int dp [|B(p)|^{-2}]^{\prime},
\end{eqnarray}
where the prime indicates the differentiation with respect to $p$.
Since the shift of the expectation value must be real valued, we can determine the integration region which satisfies
\begin{eqnarray}
\label{condition}
\int dp [|B(p)|^{-2}]^{\prime}=0,
\end{eqnarray}
provided that $C\neq0$.
From the periodicity of $B(p)$ in Eq. \Ref{B_def}, we can adopt $-\pi/2g \leq p \leq \pi/2g$ as the integration region.
From Eqs. \Ref{C_0} and \Ref{intF1}, the normalization constant $C$ becomes 
\begin{eqnarray}
\label{C}
|C|^{2} =\frac{g|\Re A_{w}|}{\pi},
\end{eqnarray}
which should not vanish.
We assume $\Re A_{w}\neq 0$ here and below. 
Then with Eqs. \Ref{qi} and \Ref{intF2}, the shift $\Delta \av{\hat{q}}$ becomes
\begin{eqnarray}
\label{Delta}
\Delta \av{\hat{q}}=\av{\hat{q}}_{f}-\av{\hat{q}}_{i}=\frac{g(|A_{w}|^{2}+1)}{2\Re A_{w}}.
\end{eqnarray}

Finally, we check that the probe wave function \Ref{xi} can realize $\av{\hat{q}}_{i}=0$ as alluded to before.
The periodicity $B^{\prime}(p+2\pi/g)/B(p+2\pi/g)=B^{\prime}(p)/B(p)$ implies that
\begin{eqnarray}
|\xi_{i}(-\pi/2g)|^{2}=|\xi_{i}(\pi/2g)|^{2},
\end{eqnarray}
and therefore
\begin{eqnarray}
\label{k}
\xi_{i}(-\pi/2g)=e^{-i\pi k/g}\xi_{i}(\pi/2g),
\end{eqnarray}
where $k$ is an arbitrary real constant. 
By choosing $k=\Delta \av{\hat{q}}$, we see that $\av{\hat{q}}_{i}=0$ and $\av{\hat{q}}_{f} =k=\Delta \av{\hat{q}}$ since Eqs. \Ref{xi} and \Ref{k} hold.

Thus, we have derived the optimal probe wave function in the momentum space as Eq. \Ref{optxi} from Eqs. \Ref{xi}, \Ref{C}, and \Ref{Delta} when $\hat{A}^{2}=1$, $\Re A_{w} \neq 0$, and the support of the function is $-\pi/2g \leq p \leq \pi/2g$.
The optimal probe wavefunction gives the maximum shift of the expectation value of the probe position as Eq. \Ref{max_shift}.

%%%%%%%%%%%%%%%%%%%%%%%%%%%%%%%%%%
\section{Summary and Discussion} 
%%%%%%%%%%%%%%%%%%%%%%%%%%%%%%%%
\label{conc_sec}

In this paper, we have derived the optimal probe wave function for the signal amplification from the weak measurement. 
The wave function in the momentum space is described as Eq. \Ref{optxi} when an observable $\hat{A}$ of the system satisfies $\hat{A}^{2}=1$ and $\Re A_{w}\neq0$, and the support of the probe wave function is $-\pi/2g \leq p \leq \pi/2g$. 
The weak measurement with the optimal probe wave function gives the maximum shift of the expectation value of the probe position as Eq. \Ref{max_shift}. which has no upper bound as $|A_{w}|$ becomes large.  The signal is sharp when we choose the optimal probe wavefunction for the weak measurement.

A few remarks are in order.
The weak measurement with the optimal probe wave function amplifies the signal more than the projective measurement. 
While our result \Ref{optxi} is restricted to the region $-\pi/2g \leq p \leq \pi/2g$, we can extend the region to $-\pi m/2g \leq p \leq \pi m/2g\ (m\in \mathbb{N})$. 
This case gives the same maximum shift and the same sharpness around the final probe position.
It is remarked that in the case of a sufficiently small coupling constant $g\ll 1$, the support of the function almost encompasses the whole momentum space.
Then, the final probe wave function in the position space behaves like the $\delta$ function.
For $\Re A_{w}=0$, the stationary solution \Ref{xi} is not normalizable, so we have excluded that case.

Practically, the wave function can be engineered by using the coupling constant $g$ and the weak value $A_{w}$ calculated from the experimental setup. 
We choose the pre- and postselected states in a given experimental setup. The value of the coupling constant needed for the construction of the wave function is initially  chosen by a reasonable guess. 
We prepare the optimal probe wave function for the chosen coupling constant. 
From the discrepancy of the theoretical prediction and the experimental data, the value of the coupling constant is narrowed down to a more precise value by iteration. 
While we showed that the signal is sharp, noise analysis is needed for the actual experiments. 
To experimentally demonstrate the optimal probe wavefunction, the spatial phase and amplitude modulation for light are needed. 
Even in with current technology, the simultaneously spatial phase and amplitude modulation has been done~\cite{Ando}. 
Therefore, in the near future, it may be possible to realize the optimal probe wavefunction.
While we have analytically shown the unbounded signal amplification, another practical approach would be to improve the signal-to-noise ratio by using the finitely amplified signal~\cite{Feizpour, Nishizawa}.

%%%%%%%%%%%%%%%%
\section*{Acknowledgments}
%%%%%%%%%%%%%%%%%
The authors acknowledge Kouji Nakamura, Atsushi Nishizawa, Masa-Katsu Fujimoto, Kentaro Somiya, and Yanbei Chen for valuable discussions and comments.
The authors are supported by the Global Center of Excellence Program ``Nanoscience and Quantum Physics" at Tokyo Institute of Technology. 
One of the authors (Y. Sh.) is also supported by JSPS (Grant No. 21008624).
%%%%%%%%%%%%%%%%%%%%%%%%%%%%%%%%
\appendix
\begin{widetext}
\section{Calculation formulas}
%%%%%%%%%%%%%%%%
\label{cal}
When $B(p)=\cos gp-iA_{w}\sin gp$, $\Re A_{w}\neq0$, and  $-\pi/2g \leq p \leq \pi/2g$, we have used the following formulas in Sec. \ref{derivation_sec}.
To calculate Eq. \Ref{qi}, we differentiate $|B(p)|^{2}$ with respect to $p$ as
\begin{eqnarray}
\label{BBprime}
[|B(p)|^{2}]^{\prime} 
&=&B^{\ast \prime}(p)B(p)+B^{\ast}(p)B^{\prime}(p) =(-g\sin gp+igA^{\ast}_{w}\cos gp)(\cos gp-iA_{w}\sin gp) +B^{\ast}(p)B^{\prime}(p) \nonumber \\
&=&-g\sin gp \cos gp +igA^{\ast}_{w} (1-\sin^{2}gp) +igA_{w}(1-\cos^{2} gp) +g|A_{w}|^{2}\sin gp \cos gp +B^{\ast}(p)B^{\prime}(p) \nonumber \\
&=&2ig\Re A_{w} +(\cos gp+iA^{\ast}_{w}\sin gp)(-g\sin gp-igA_{w}\cos gp)+B^{\ast}(p)B^{\prime}(p) \nonumber \\
&=&2ig\Re A_{w}+2B^{\ast}(p)B^{\prime}(p).
\end{eqnarray}
Then, to calculate Eqs. \Ref{C} and \Ref{Delta}, respectively, we have used two integration formulas:
\begin{eqnarray}
\label{intF1}
|C|^{-2}=\int_{-\frac{\pi}{2g}}^{\frac{\pi}{2g}} dp |B(p)|^{-2}&=&\int_{-\frac{\pi}{2g}}^{\frac{\pi}{2g}} \frac{dp}{\cos^{2} gp}\frac{1}{1+2\Im A_{w}\tan gp+|A_{w}|^{2}\tan^{2}gp} =\int_{-\infty}^{\infty} \frac{dx}{g}\frac{1}{1+2\Im A_{w}x+|A_{w}|^{2}x^{2}}  \nonumber \\
&=&\frac{|A_{w}|^{2}}{g(\Re A_{w})^{2}} \int_{-\infty}^{\infty} dx\frac{1}{1+\left[ (\Im A_{w}+|A_{w}|^{2}x)/\Re A_{w} \right]^{2}} =\frac{\pi}{g|\Re A_{w}|}
\end{eqnarray}
and
\begin{eqnarray}
\label{intF2}
\int_{-\frac{\pi}{2g}}^{\frac{\pi}{2g}} dp |B(p)|^{-4}
&=&\int_{-\frac{\pi}{2g}}^{\frac{\pi}{2g}} \frac{dp}{\cos^{4} gp} \frac{1}{(1+2\Im A_{w}\tan gp+|A_{w}|^{2}\tan^{2}gp)^{2}} \nonumber \\
&=&\frac{1}{g}\int_{-\infty}^{\infty} dx \frac{1+x^{2}}{(1+2\Im A_{w}x+|A_{w}|^{2}x^{2})^{2}} =\frac{|A_{w}|^{2}+1}{2(\Re A_{w})^{2}}\int_{-\infty}^{\infty} \frac{dx}{g} \frac{1}{1+2\Im A_{w}x+|A_{w}|^{2}x^{2}} \nonumber  \\
&&+\frac{1}{2g(\Re A_{w})^{2}|A_{w}|^{2}}\int_{-\infty}^{\infty} dx \left( \frac{\Im A_{w} (|A_{w}|^{2}+1)+(|A_{w}|^{4}+|A_{w}|^{2}-2(\Re A_{w})^{2})x}{1+2\Im A_{w}x+|A_{w}|^{2}x^{2}}\right)^{\prime} \nonumber  \\
&=&\frac{|A_{w}|^{2}+1}{2(\Re A_{w})^{2}}\int_{-\frac{\pi}{2g}}^{\frac{\pi}{2g}} dp |B(p)|^{-2} =\frac{|A_{w}|^{2}+1}{2(\Re A_{w})^{2}}|C|^{-2}.
\end{eqnarray}
In these derivations, we have used the substitution $x=\tan gp$.\\
\end{widetext}
%%%%%%%%%%%%%%%%
\section{The weak value in the Mach-Zehnder interferometer}
%%%%%%%%%%%%%%%%
\label{M_Z}
We give the state evolutions and the weak value in the experimental setup shown in Fig. \ref{Mach}. 
The initial polarization state $\ket{\Pi}=a\ket{H}+b\ket{V}$ is separated into the horizontal state $\ket{H}$ and the vertical state $\ket{V}$ by the PBS.
Assuming that state $\ket{H}$ goes through to path B and the vertical state $\ket{V}$ is reflected to path C, we can express the preselected state as
\begin{eqnarray}
\ket{\phi_{i}}=a\ket{H}\otimes\ket{B} + b \ket{V}\otimes\ket{C}.
\end{eqnarray}
The intermediate state, which is the state in ports D and E after the BS before postselection by the polarizer, is described as 
\begin{eqnarray}
\ket{\phi_{m}}&=&\frac{1}{\sqrt{2}}(-a\ket{H}+b\ket{V})\otimes\ket{D} \nonumber \\
&&+\frac{1}{\sqrt{2}}(a\ket{H}+b\ket{V})\otimes\ket{E}.
\end{eqnarray}
We note that the polarization state in port E coincides with the initial state $\ket{\Pi}$.
The postselected state is given by the tunable angle $\varphi$ of the polarizer:
\begin{eqnarray}
\ket{\phi_{f}}=\frac{1}{\sqrt{2}}(\cos \varphi \ket{H}+ \sin \varphi \ket{V})\otimes\ket{D}.
\end{eqnarray}
The weak value of the projection operator $\kb{C}$ to path C for the observable of the system is calculated as
\begin{eqnarray}
C_{w}=\frac{\bkt{\phi_{f}}{C}\bkt{C}{\phi_{i}}}{\bkt{\phi_{f}}{\phi_{i}}}=\frac{b\sin \varphi}{-a\cos \varphi+b\sin\varphi}.
\end{eqnarray}
For $a=\cos \chi$ and $b=\sin \chi$, the weak value becomes
\begin{eqnarray}
C_{w}=-\frac{\sin \chi\sin \varphi}{\cos(\chi + \varphi)}.
\end{eqnarray}
We easily see that when $\chi + \varphi$ approaches $\pi/2$, the absolute value of $C_{w}$ gets arbitrarily large.
When we set $\hat{A}=2\kb{C}-1$, we get $\hat{A}^{2}=1$.

%%%%%%%%%%%%%%%%%%%%%%%%%%%%%%%%


\begin{thebibliography}{99}
\bibitem{Lebesgue}
In the general smoothing cases of the optimal probe wave function \Ref{optxi}, the probe wave functions in the position space (\ref{initial_probe_position}) and (\ref{final_probe_position}) 
do not drastically change as guaranteed by Lebesgue's dominated convergence theorem.
\end{thebibliography}
\end{document}